\documentclass[3p,times,twocolumn]{elsarticle}
\pdfoutput=1

\usepackage[english]{babel}

\usepackage{amsmath}

\usepackage{amssymb}

\usepackage[figuresright]{rotating}
\usepackage{subcaption}

\journal{arXiv}

\begin{document}

\begin{frontmatter}




\title{Search for Heavy Right Handed Neutrinos at the FCC-ee}


\author[ge]{A. Blondel (presenter)}
\author[zh]{E. Graverini}
\author[zh]{N. Serra}
\author[epfl]{M. Shaposhnikov}

\address[ge]{DPNC, University of Geneva, Quai Ansermet 24, CH-1205 Geneva, Switzerland}
\address[zh]{Physik Institut, University of Zurich, CH-8057 Zurich, Switzerland}
\address[epfl]{ITPP, EPFL, CH-1015 Lausanne, Switzerland}

\begin{abstract}
The Standard Model of particle physics is still lacking an understanding of the generation and nature of neutrino masses. A favorite theoretical scenario (the \textit{see-saw} mechanism) is that both Dirac and Majorana mass terms are present, leading to the existence of heavy partners of the light neutrinos, presumably massive and nearly sterile. These heavy neutrinos can be searched for at high energy lepton colliders of very high luminosity, such as the Future electron-positron $e^+e^-$ Circular Collider, FCC-ee (TLEP), presently studied within the Future Circular Collider design study at CERN, as a possible first step. A first look at sensitivities, both from neutrino counting and from direct search for heavy neutrino decay, are presented. The number of neutrinos $N_\nu$ should be measurable with a precision of  $\Delta N_\nu \approx \pm(0.0004 - 0.0010) $, while the direct search appears very promising due to the long lifetime of heavy neutrinos for small mixing angles. A sensitivity down to a heavy-light mixing of ${|U|}^2 \simeq 10^{-12}$ is obtained, covering a large phase space for heavy neutrino masses between 10 and 80 GeV/$c^2$.

\end{abstract}


\end{frontmatter}


\section{Introduction}\label{sec:Intro}
\begin{figure}
	\centering
	\includegraphics[width=8.cm]{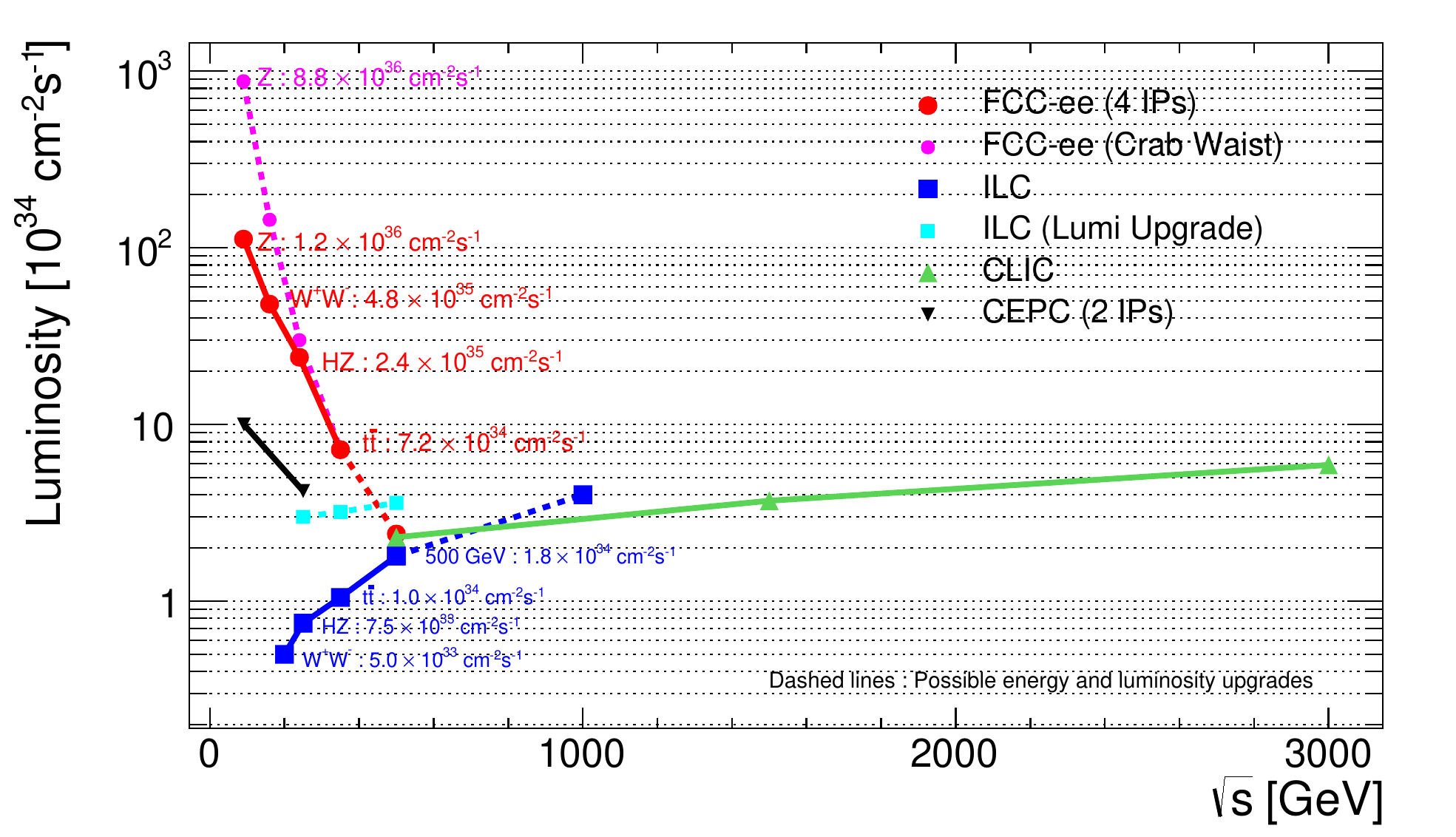}
	\caption{Foreseen FCC-ee, ILC and CLIC luminosities as a function of the center-of-mass energy. Also shown are the ILC luminosity upgrade points implying a reconfiguration of the RF power sources for low energy running~\cite{Harrison:2013nva}, and the FCC-ee possible improvement using the crab-waist scheme~\cite{PhysRevSTAB.17.041004}.}\label{fig:lumi}
\end{figure}

Following the discovery of the Higgs Boson~\cite{Aad:2012tfa,Chatrchyan:2012ufa}, there has been growing interest in high-luminosity $e^+e^-$ colliders aimed at the detailed study of this particle of a new nature, especially since they could be followed by a high-energy pp collider in the same tunnel~\cite{Blondel:2011fua,Blondel:2012ey,Blondel:2013rn,Koratzinos:2013ncw,Koratzinos:2013chw,Gomez-Ceballos:2013zzn}. Recently a design study has been launched of the accelerators that could fit in a $\sim$100 km tunnel, the Future Circular Collider design study, including a high luminosity $e^+e^-$ storage ring collider FCC-ee, able to address center-of-mass energies between 90 and 350~GeV, as a possible first step, and with a 100 TeV proton collider as ultimate goal~\cite{fcc-design,fccee-design}. The use of techniques inspired from the b-factory technique, a low-beta ring coupled with a full-energy top-up booster, allows the FCC-ee to achieve very high luminosities, which distinguishes it from the Linear Collider that is, however, able to reach higher energies. The foreseen luminosity, taking into account the capability of circular colliders to provide collisions at up to four interaction points, is shown on Figure~\ref{fig:lumi}. These curves very much encompass the different physics cases for linear and circular high energy lepton colliders. What is notable is the very high luminosity of the FCC-ee at the Z pole, allowing a total of $10^{12}$ $Z$ bosons -- or $10^{13}$ with the crab-waist scheme~\cite{PhysRevSTAB.17.041004} -- to be produced in a few years. This is the key to the investigation of extremely rare decays. This contribution describes the example of the search for sterile right-handed partners of neutrinos that were so far assumed to be completely out of reach for high energy colliders.

\section{Right handed neutrinos}
\begin{figure}
	\centering
	\begin{subfigure}[]{\linewidth}
		\includegraphics[width=8.cm]{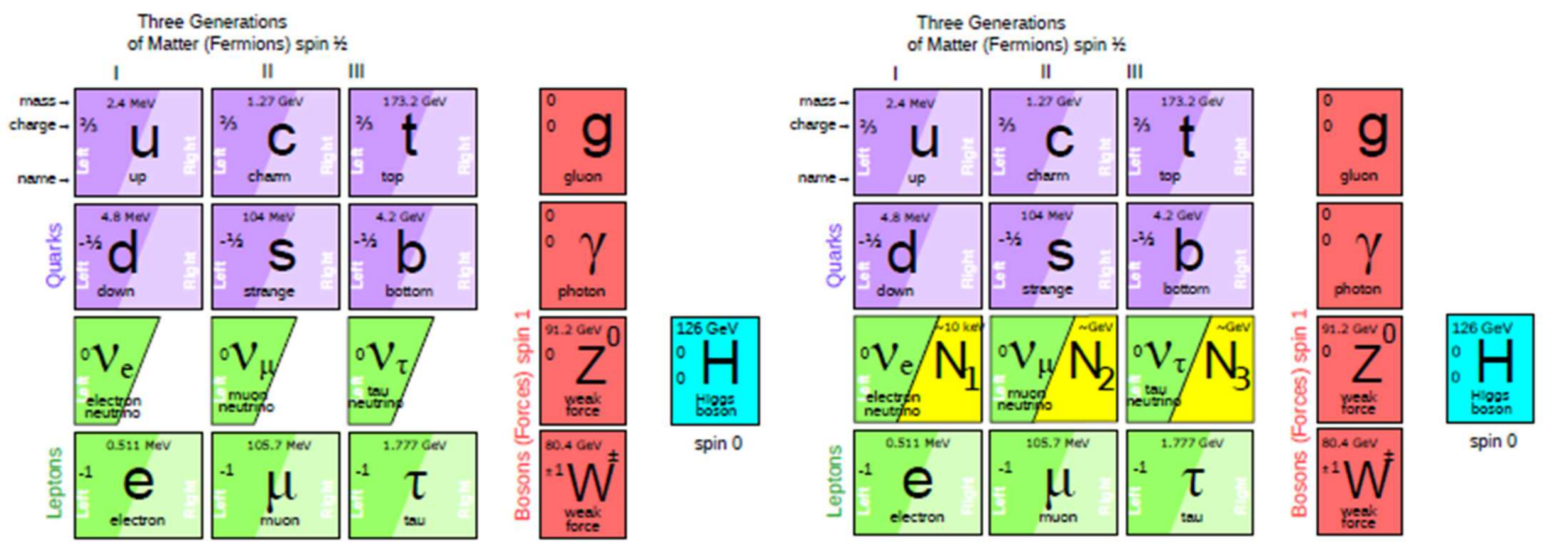}
	\end{subfigure}
	\begin{subfigure}[]{\linewidth}
		\vspace{0.4cm}
		\includegraphics[width=8.cm]{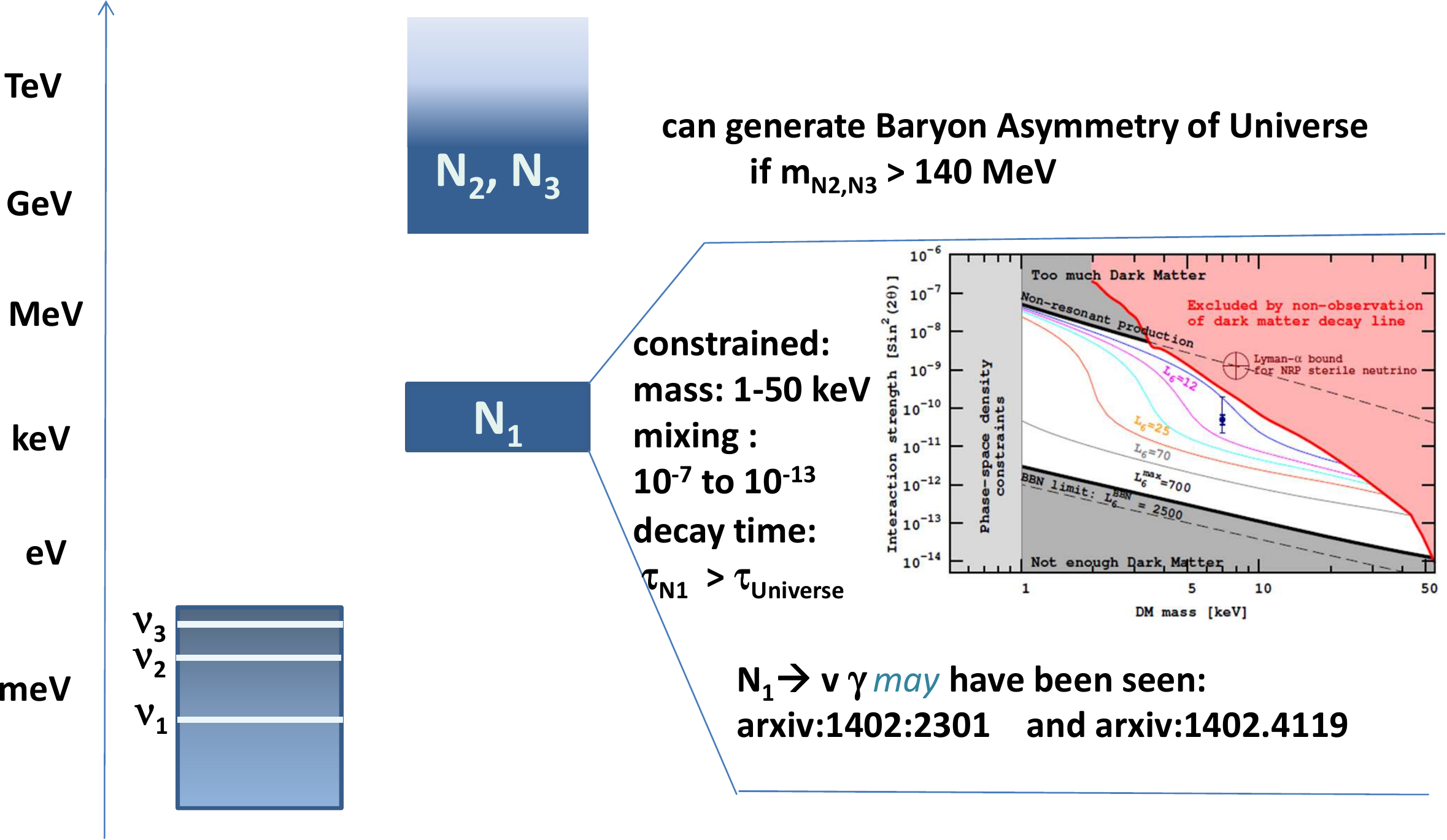}
	\end{subfigure}
	\caption{Top left: the Standard Model, incomplete without two chiralities for each massive neutrino; top right: a possible scenario. Bottom: a possible scenario in which right-handed neutrinos provide a dark matter candidate and the baryon asymmetry of the Universe.}\label{fig:scenario}
\end{figure}

Neutrino experiments have provided compelling evidences for the existence of neutrino oscillations, caused by nonzero neutrino masses and neutrino mixing, in contrast with the predictions of the Standard Model (SM)~\cite{rev14}. Not only does the SM assume massless neutrinos, but there is no unique way to introduce neutrino masses in it. A Dirac mass term as for the other fermions seems natural, and would imply the existence of right handed neutrinos. A Majorana mass term is not forbidden for neutrinos because they are neutral. The coexistence of both terms would lead to the (type I) see-saw mechanism, first introduced in the context of Grand Unified Theories~\cite{Minkowski:1977sc,Yanagida:1979as,GellMann:1980vs,Mohapatra:1979ia}. In either case the existence of right-handed partners to the well-known (left-handed) neutrinos is predicted. In the Grand Unified Models, the right-handed neutrinos typically get masses of the order of $10^{10}$ GeV, which, in combination with Dirac masses of the same magnitude as that of the top quark, generates the observed light neutrino masses. More recently, consistent models with Electroweak scale Majorana masses (sub-MeV to TeV region) have been proposed, suggesting Dirac masses smaller or of the same order of magnitude as the electron mass~\cite{Asaka:2005an}. These models are of considerable interest, since the three families of right-handed neutrinos allow generation of the Baryon asymmetry of the Universe (BAU) \cite{Akhmedov:1998qx,Asaka:2005pn}, and, for a suitable parameter space for the lightest of the heavy right-handed neutrinos, could even produce a suitable dark matter candidate~(for a review see \cite{Boyarsky:2009ix}, Figure~\ref{fig:scenario}). The signature for such heavy neutrino dark matter candidate is its decay $N_1 \to \nu \gamma$. The constraints of abundance of dark matter and the absence of detection of the decay lead to the constraint shown in Figure~\ref{fig:scenario}. A possible indication of the decay has been reported in~\cite{Bulbul:2014sua,Boyarsky:2014jta}, but this needs to be confirmed. 

\section{Phenomenology of right handed neutrino production and decay}
The phenomenology of the production of Right-Handed neutrinos has been described already in~\cite{Gronau:1984ct,Gorbunov:2007ak,Atre:2009rg,Cely:2012bz,Keung:1983uu}. Right-handed neutrinos have both weak isospin and charge equal to zero, and have no other interaction than Yukawa couplings to the Higgs boson and leptons. For this reason they are often called ``sterile''. In the see-saw picture, the diagonalization of the mass matrix with two mass terms per neutrino family leads to a light, almost left-handed neutrinos (the known one) and a heavier, almost right-handed neutrinos -- thus they are often called ``Heavy Neutral Leptons''; in this picture both the right-handed and left-handed neutrinos are Majorana particles, and neutrino-less beta-decay is possible. The heavy  neutrinos only interact via their mixing to the light ones.

Let us first consider the situation for one family of neutrinos. The diagonalization of the neutrino mass-matrix with both Dirac $m_D$, and Majorana $M$ mass terms leads to new mass eigenstates, that we will note $\nu$ and $N$ respectively and relate to the chirality eigenstates by a mixing angle $\theta$, such that $\theta^2 \approx m_\nu / m_N \approx {(m_D/M)}^2$, with $m_N \approx M$, and $m_\nu \approx m_D^2/M$.

Thus the mass spectrum defines completely the mixing angle. The mass of the light neutrino is unknown and we assume for the discussion that it is bound between the cosmological limit of about 0.2~eV and the lower limit given by the square root of the measured oscillation mass differences $\Delta m_{12}^2 = (7.58\pm 0.24) \times 10^{-5}$ eV$^2$ and $|\Delta m_{23}^2| = (2.35 \pm 0.12) \times 10^{-5}$ eV$^2$. The situation is more complicated with several generations of HNLs.

\begin{figure}
	\centering
	\begin{subfigure}[]{\linewidth}
		\includegraphics[width=8.cm]{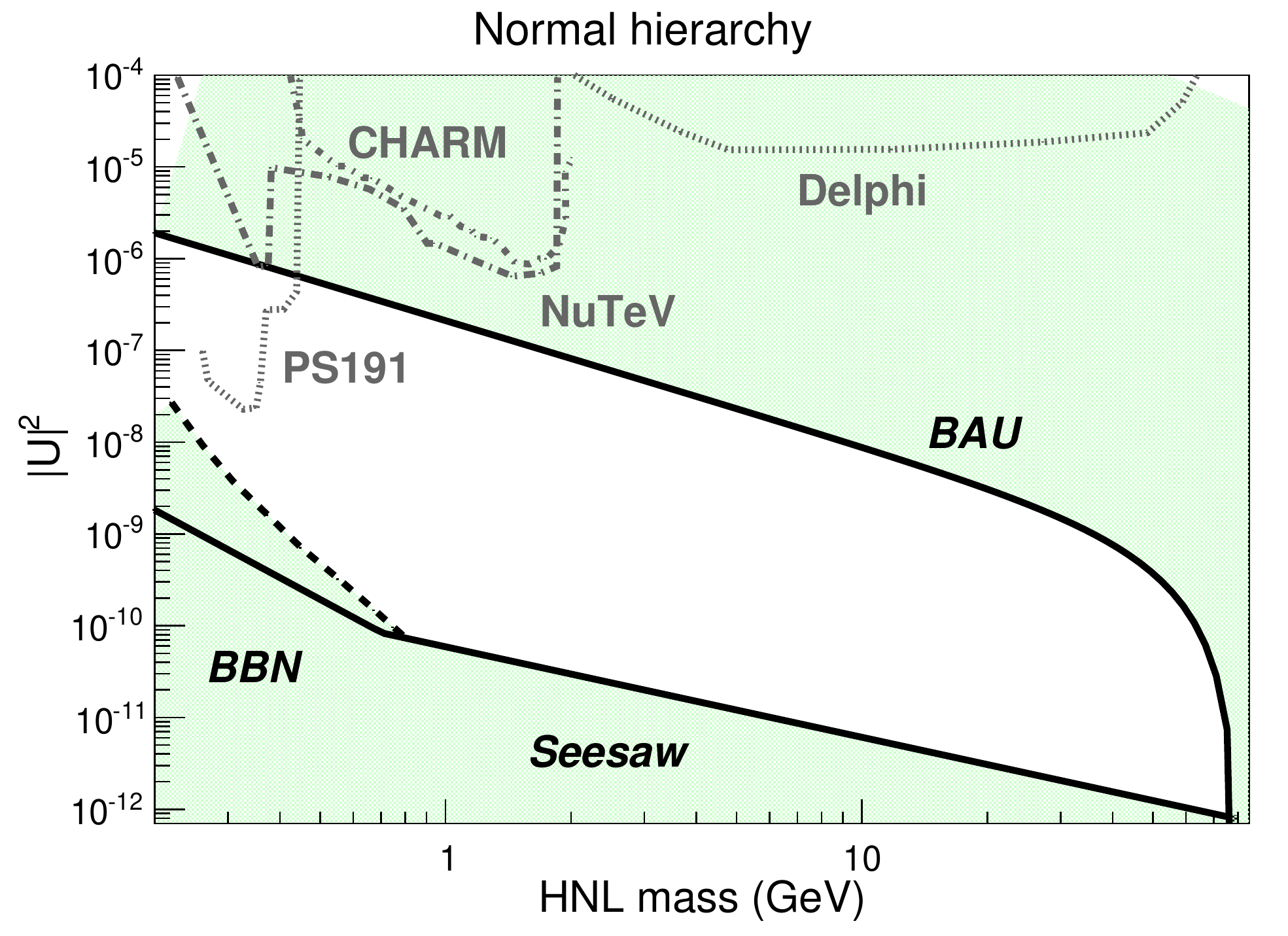}
	\end{subfigure}
	\begin{subfigure}[]{\linewidth}
		\includegraphics[width=8.cm]{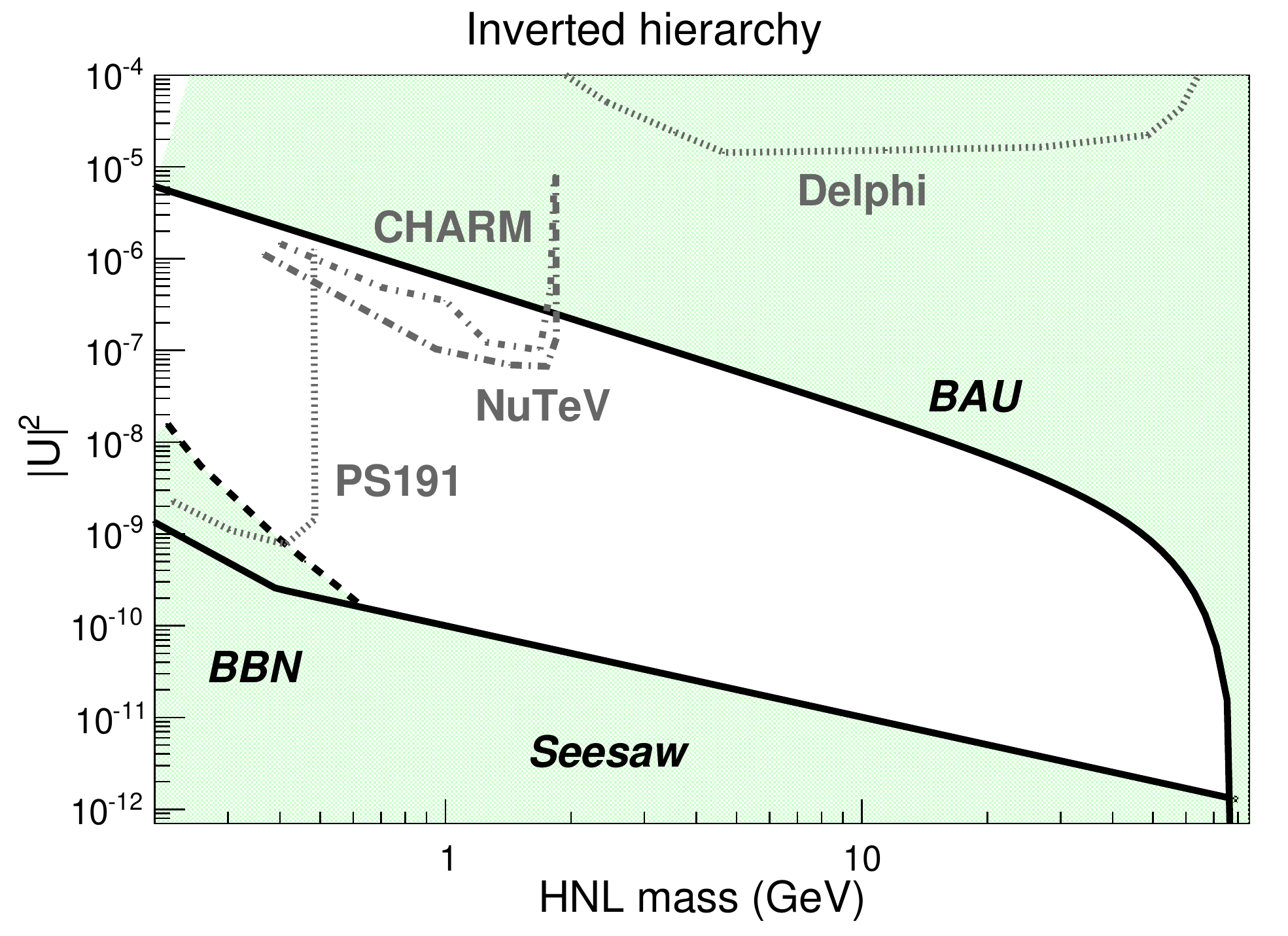}
	\end{subfigure}
	\caption{Interesting domains in the mass-coupling parameter space of heavy neutrinos and current experimental limits, for normal and inverted hierarchy of the left-handed neutrino masses.}\label{fig:parspace}
\end{figure}

A three family analysis of these constraints for right-handed neutrinos with masses below 10~GeV is found in~\cite{Canetti:2012kh}. In Figure~\ref{fig:parspace} we extend the range up to the mass of the intermediate vector boson $W$. The see-saw line gives a lower limit on the mixing angle of right-handed neutrinos with active neutrinos. Below this line, the active neutrino mass differences observed in neutrino experiments cannot be accounted for in the GeV scale see-saw mechanism. Above the BAU  line the reactions with right-handed neutrinos are in thermal equilibrium during the relevant period of the Universe expansion, making the baryogenesis due to right-handed neutrino oscillations impossible. For $m_N$ close to $M_W$ and above $M_W$ the rate of reactions with $N$'s is enhanced due to the kinematically allowed decay $N \to \ell W$ leading to stronger constrains on the mixing \cite{Shaposhnikov:2008pf}.  The BAU curve intersects with the see-saw line at $m_N = M_W$, so that the parameter space is bound on all sides. 

For even larger masses of $N$ another mechanism of baryogenesis -- resonant leptogenesis -- can operate~\cite{Pilaftsis:2003gt}. This part of the parameter space cannot be directly studied with the FCC-ee operated at the $Z$ resonance.

The production and decay of the heavy neutrino in $Z$ decays has already been studied at LEP by the L3 and DELPHI collaborations~\cite{Abreu:1996pa,Adriani:1992pq}. It is largely determined by the mixing angle. When a left-handed neutrino is produced e.g. in $Z$ decays it is actually a mixture of the light and heavy state: 
\begin{equation*}
	\nu_L = \nu \cos\theta + N\sin\theta
\end{equation*}
with $\theta^2 \approx m_\nu / m_N$. Thus the decay width of the $Z$ into a pair of light and heavy neutrinos will be given by 
\begin{equation*}
	\Gamma_{Z\to\nu N} = 3 \Gamma_{Z\to\nu\nu}^{SM} {|U|}^2 {\left(1 - \left(m_N/m_Z\right)^2\right)}^2 \left(1 + \left(m_N/m_Z\right)^2\right)
\end{equation*}
with ${|U|}^2 \sim \theta^2$. The best existing limits are around ${|U|}^2 < 10^{-5}$ in the mass range relevant to high energy investigations (Figure~\ref{fig:parspace}). The mixing of sterile neutrinos with the active neutrinos of each flavour $i$ is defined as ${|U_i|}^2$, where $i=e,\mu,\tau$. The total mixing ${|U|}^2$ is defined as ${|U|}^2 = \sum_i {|U_i|}^2$. The measurement of the partial width is sensitive to ${|U|}^2$, while in direct searches the final state depends on the relative strength of the partial ${|U_i|}^2$. In our analysis we consider the combination of ${|U_i|}^2$ allowed by present constrains from neutrino oscillations that maximises the BAU.

\begin{figure}
	\centering
	\includegraphics[width=6.5cm]{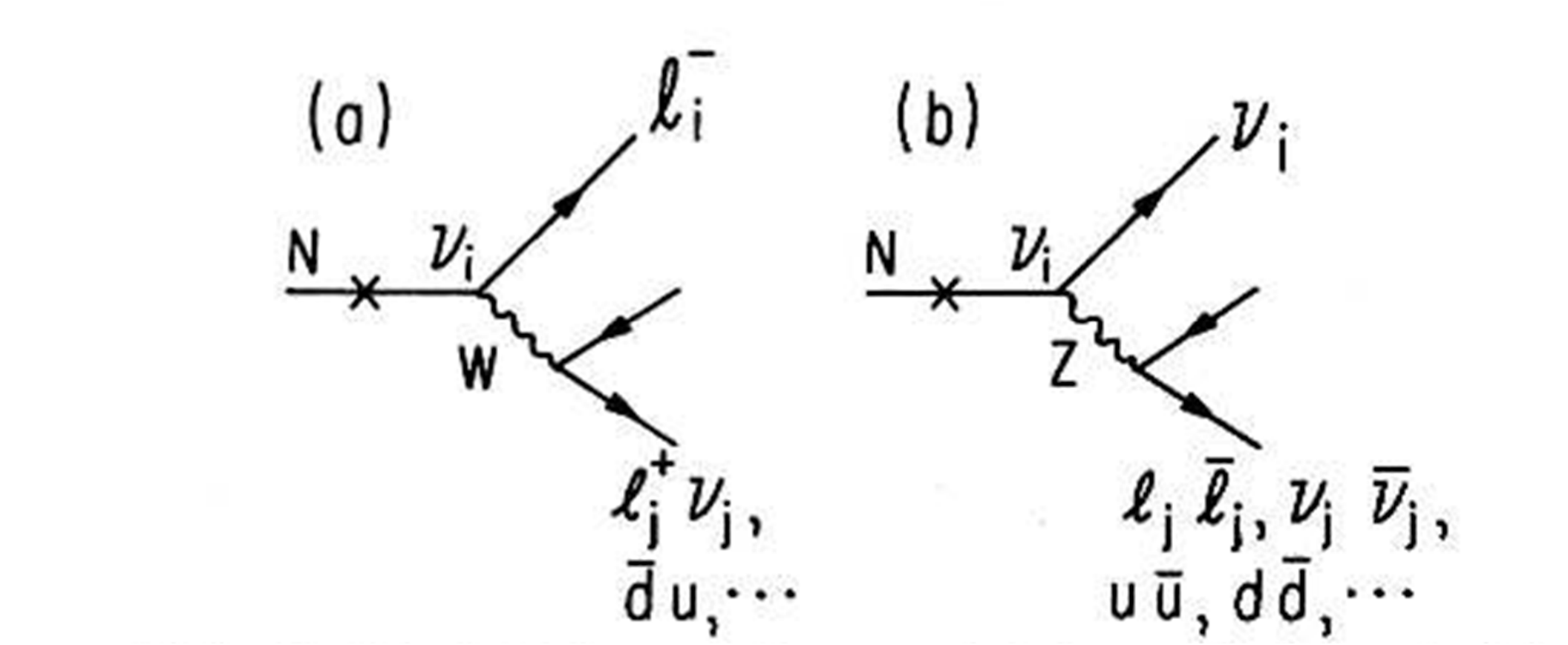}
	\caption{Decay modes of heavy neutrinos through mixing with light neutrinos: the charged current decay $N \to \ell \nu$ (a), the neutral current decay $N \to \nu + \gamma/Z$.}\label{fig:feyn}
\end{figure}

Heavy neutrinos decay as shown in Figure~\ref{fig:feyn}. At large masses the fully visible decay $N \to \ell W (W\to qq)$ accounts for more than 50\% of the decays. 

The decay rate of the heavy neutrino depends very strongly on the mass, both via the three body phase space (in the fifth power of mass) but also through the mixing angle. The average decay length is given in~\cite{Abreu:1996pa}:
\begin{equation*}
	L\sim \frac{3[\text{cm}]}{{|U|}^2 \times {(m_N [\text{GeV}])}^6}
\end{equation*}

The existence of heavy neutrinos in the accessible mass range would manifest itself in many different ways in high energy colliders. 

\begin{figure}
	\centering
	\includegraphics[width=8.cm]{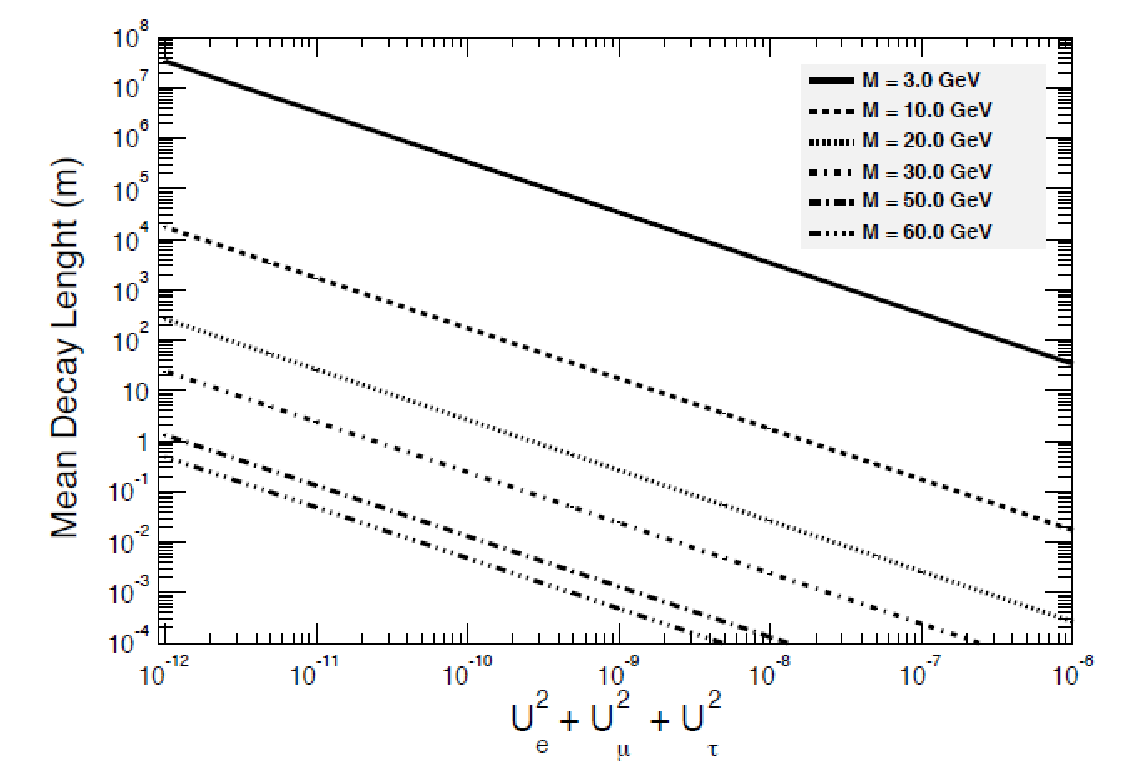}
	\caption{Mean decay length of a heavy right-handed neutrino of mass $M$ as a function of the sum of its couplings to the three families of light neutrinos.}\label{fig:ctau}
\end{figure}

If $N$ is heavy but within kinematical reach, it will decay in the detector and could be detected and possibly discovered. If not, the suppression factor due to kinematics or the mixing angle to a kinematically forbidden state could lead to an apparent  PMNS matrix unitarity violation and deficit in the $Z$ ``invisible'' width, as explained in~\cite{Jarlskog:1990kt}. It could also appear in $c$, $b$, $W$ and top decays as exotic decays or apparent violations of the SM predictions, or generate invisible or exotic Higgs decays. 

The lifetime of heavy heutrinos produced in $Z$ decays is shown in Figure~\ref{fig:ctau}. Note that, for a mass of 50~GeV and light neutrino mass of 0.05 eV, ${|U|}^2\sim 10^{-12}$ and the decay length is about one meter. 

In the following we concentrate on $Z$ decays.

\section{The $Z$ invisible width measurement}
\begin{figure}
	\centering
	\includegraphics[width=6.cm]{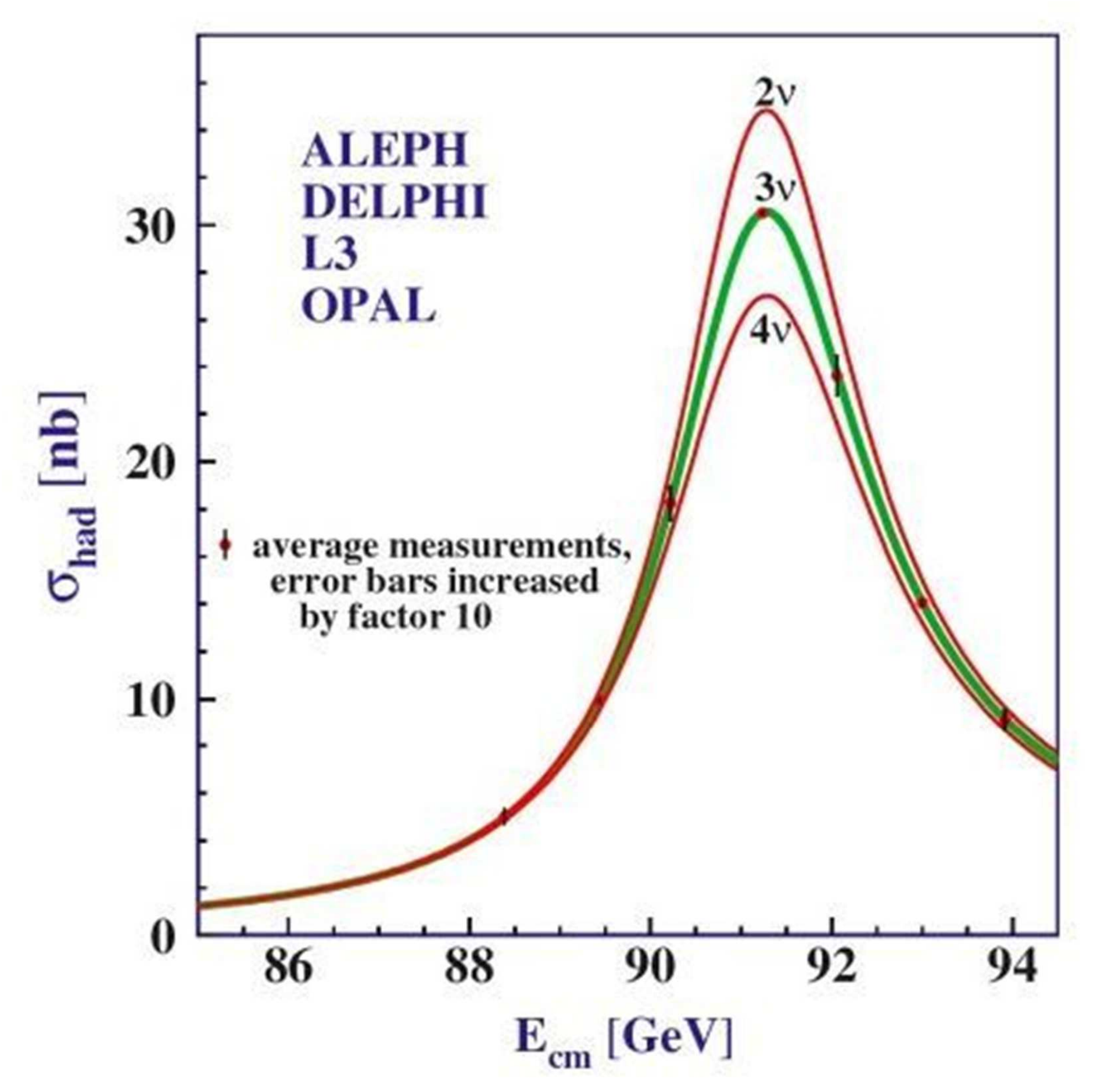}
	\caption{The measurement of the number of neutrino families at LEP, mostly from the hadronic cross-section at the $Z$ peak.}\label{fig:nnu}
\end{figure}

The measurement of the number of neutrino species at LEP~\cite{ALEPH:2005ab} was performed using the fit of the $Z$ line shape, and was dominated by the measurement of the peak hadronic cross-section, Figure~\ref{fig:nnu}. Effectively what was measured is the ratio of the $Z$ invisible width to the leptonic partial width, expressed as number of neutrinos:
\begin{equation*}
	n_\nu \equiv {\left( \frac{\Gamma_{inv}}{\Gamma_{lept}} \right)}^{meas} \bigg/\, {\left( \frac{\Gamma_{\nu\bar{\nu}}}{\Gamma_{lept}} \right)}^{SM}
\end{equation*}
So defined, the measurement is largely insensitive to radiative corrections affecting e.g. $\alpha(m_Z)$, the $\rho$ parameter, or the hadronic partial widths of the $Z$.
Interestingly, the result
\begin{equation*}
	N_\nu = 2.9840 \pm 0.0082
\end{equation*}
\cite{ALEPH:2005ab} is nearly two standard deviations lower than 3, going in the direction of the deficit expected from sterile neutrinos. The difficulty with improving this measurement is that it was already very much systematically limited by the theoretical error on the cross-section normalization using Bhabha scattering. With new calculations, it is estimated that this measurement can be at best improved by a factor 2-3 in precision.

Another technique was proposed to measure the invisible width, using the $e^+ e^- \to Z \gamma$ radiative return process from a centre-of-mass energy higher than the $Z$ resonance. The nearly monochromatic photon provides a tag of the reaction, for which it is required that nothing else is seen in the event. This measurement was also performed at LEP yielding $N_\nu = 2.92 \pm 0.05$, strongly limited by statistics~\cite{revNnu}.

A better way to perform this measurement at a high luminosity machine such as FCC-ee was proposed in~\cite{Gomez-Ceballos:2013zzn}, namely to measure the ratio
\begin{equation*}
n_\nu \equiv {\left( \frac{ e^+e^- \to \gamma Z_{inv} }{ e^+e^- \to \gamma Z_{lept} } \right)}^{meas} \bigg/\, {\left( \frac{\Gamma_{\nu\bar{\nu}}}{\Gamma_{lept}} \right)}^{SM}
\end{equation*}
by comparing the number of times the tagging photon is accompanied with nothing compared with the times it is accompanied with a pair of leptons. The method still needs to be fully developed, but the use of the photon tag should ensure very small systematic errors. Preliminary studies~\cite{6th-fcc-ws} have shown that the method should also be safe from the point of view of the photonic radiative corrections. It has been estimated that this measurement can be made with a statistical precision of $\Delta N_\nu = \pm 0.0008$ parasitically with a run at the $W$ pair threshold of 2~years at 1.6~ab$^{-1}$/year at $E_{CM}=161$~GeV. A dedicated run of one year at $E_{CM}=105$~GeV using the crab-waist scheme would lead an error of less than $\Delta N_\nu = \pm 0.0004$, or a sensitivity of ${|U|}^2 \sim 3\times 10^{-4}$ for a sterile neutrino search.

These results are extremely important in the context where the Z invisible width can reveal dark matter candidates, as pointed out in e.g.~\cite{deSimone:2014pda}. It is clear, however, that this method cannot reach the precision required to detect sterile neutrinos with the very small mixings expected from see-saw models. 

\section{Direct search in $Z$ decays}
The direct search for sterile neutrinos in $Z$ decays consists in looking for events with one light neutrino produced in association with a heavy one, that decays according to the diagrams of Figure~\ref{fig:feyn}. This is the method already used at LEP~\cite{Abreu:1996pa,Adriani:1992pq}. The limitation comes from the four fermion processes such as $Z \to W^\star W \to \ell \nu q q$. If it were not for the lifetime of the heavy neutrino this method would be quickly limited by the background  to a sensitivity of around ${|U|}^2 \sim 10^{-6}$.

A dramatic change arises when the lifetime of the heavy neutrino is taken into account. For very small mixings that are indeed expected, the decay length shown in Figure~\ref{fig:ctau} becomes substantial, and a detached vertex topology will arise. Note that, while the neutral current decays $N \to \nu + \gamma/Z$ always feature missing neutrinos in the final state, charged current decays $N \to W \ell$ can be completely reconstructed when the $W$ decays into hadrons.

\begin{figure}
	\centering
	\includegraphics[width=5.cm]{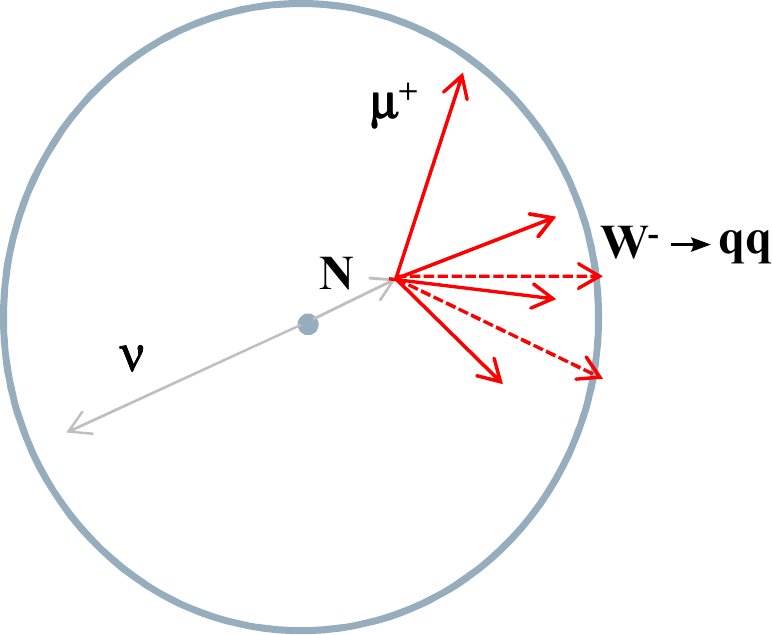}
	\caption{Sketch of the topology of a $Z \to \nu N$ decay, with $N$ subsequently decaying into $\mu^+W^-$.}
\end{figure}

It is difficult to imagine any significant background to the search for a 20-80~GeV object decaying 1~m away from the interaction point, in an $e^+e^-$ machine with no pile up. Atmospheric neutrino interactions in the detector will arise at the rate of a few tens per year, but the requirement that the observed detached particles form a vertex pointing back to the IP, with the correct mass and time-of-flight, is expected to kill backgrounds very efficiently. 

An exposure of a few years at $Z$ peak with the maximal luminosity would yield $10^{13}$ $Z$ particles, thus $2\times 10^{12}$ $Z \to \nu \bar{\nu}$ events. A mixing of ${|U|}^2 \sim 10^{-12}$ would yield a few dramatic candidates. 

\begin{figure}
	\begin{subfigure}[]{\linewidth}
		\includegraphics[width=8.0cm]{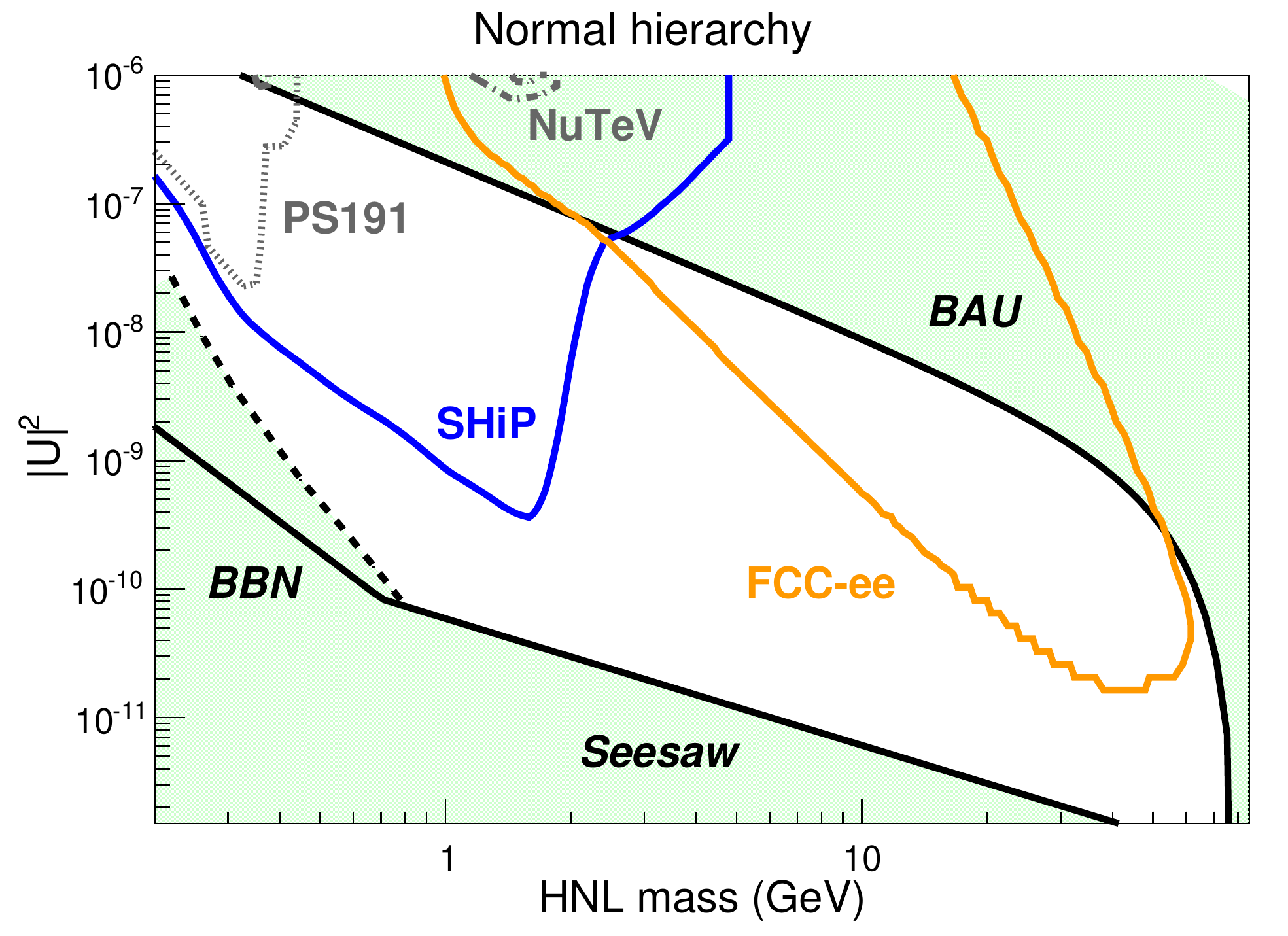}
		\caption{Decay length 10-100~cm, $10^{12}$ $Z^0$}
	\end{subfigure}
	\begin{subfigure}[]{\linewidth}
		\includegraphics[width=8.0cm]{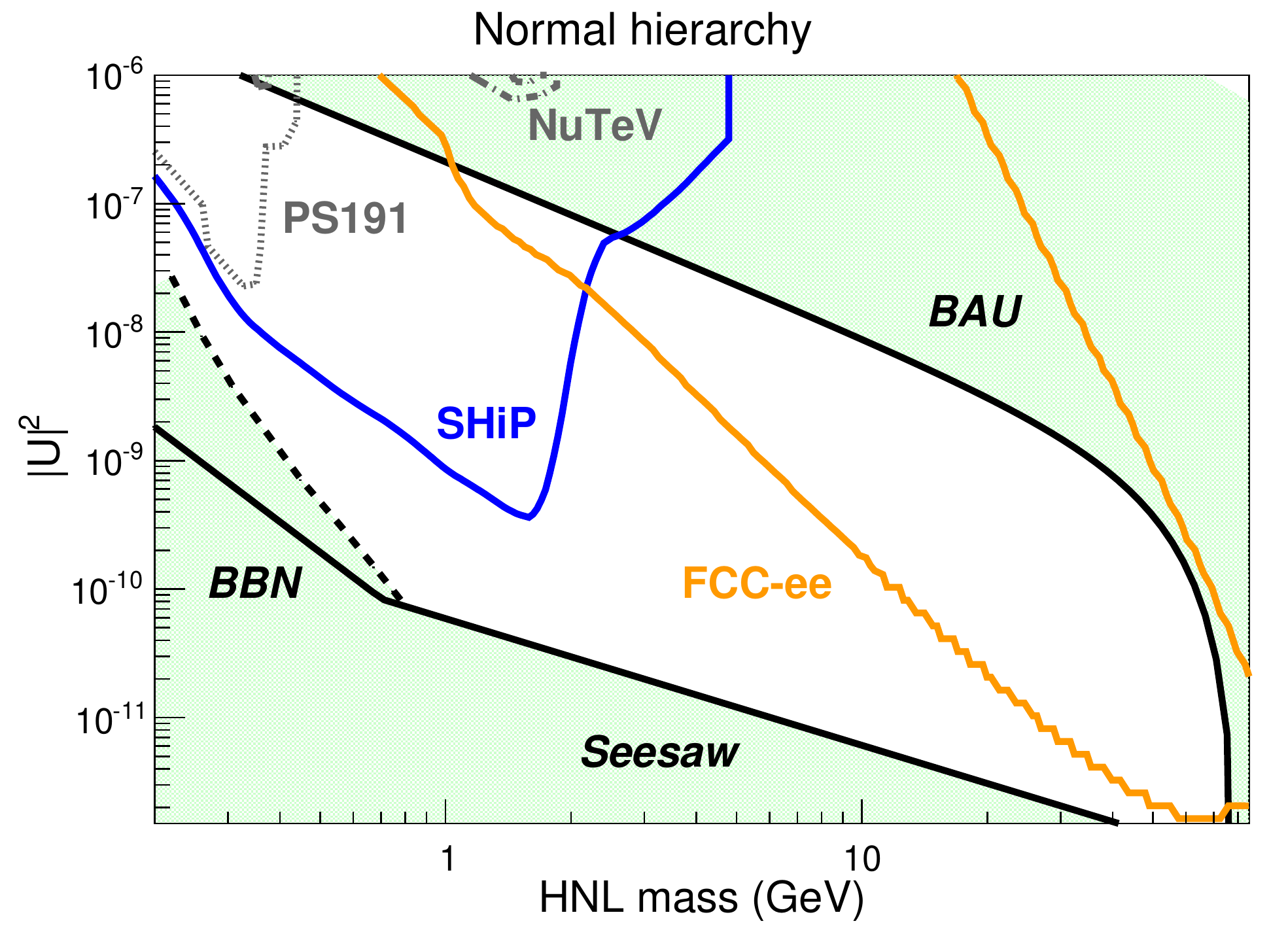}
		\caption{Decay length 10-100~cm, $10^{13}$ $Z^0$}
	\end{subfigure}
	\begin{subfigure}[]{\linewidth}
		\includegraphics[width=8.0cm]{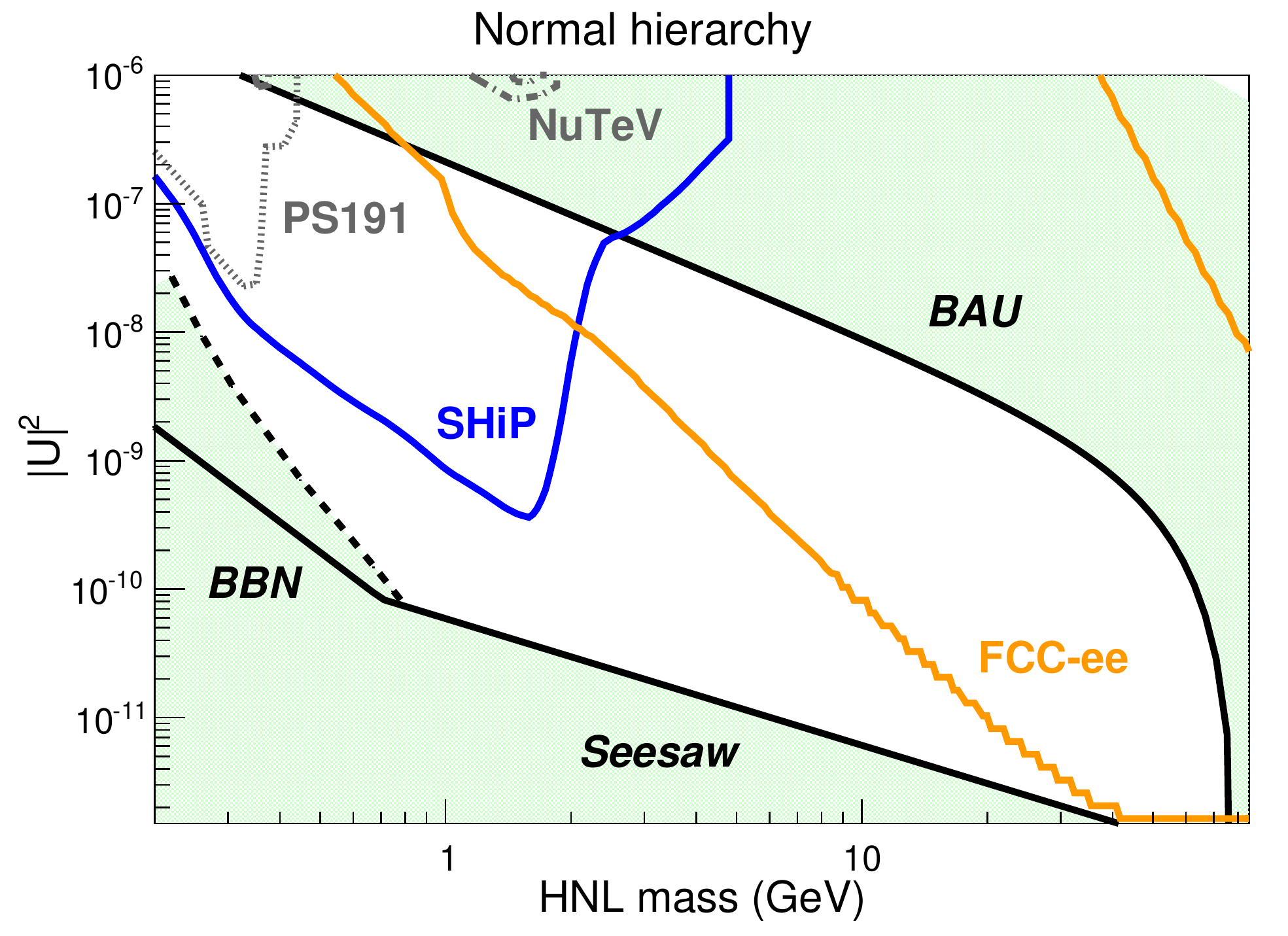}
		\caption{Decay length 0.01-500~cm, $10^{13}$ $Z^0$}
	\end{subfigure}
	\caption{Regions of sensitivity for sterile neutrinos as a function of mass and mixing to light neutrinos (normal hierarchy): for $10^{12}$ $Z$ decays occurring between 10~cm and 1~m from the interaction point (a), same for $10^{13}$ $Z$ decays (b), for $10^{13}$ $Z$ decays occurring between 100~$\mu$m and 1~m from the interaction point (c).}\label{fig:sensitivity-nh}
\end{figure}

\begin{figure}
	\begin{subfigure}[]{\linewidth}
		\includegraphics[width=8.0cm]{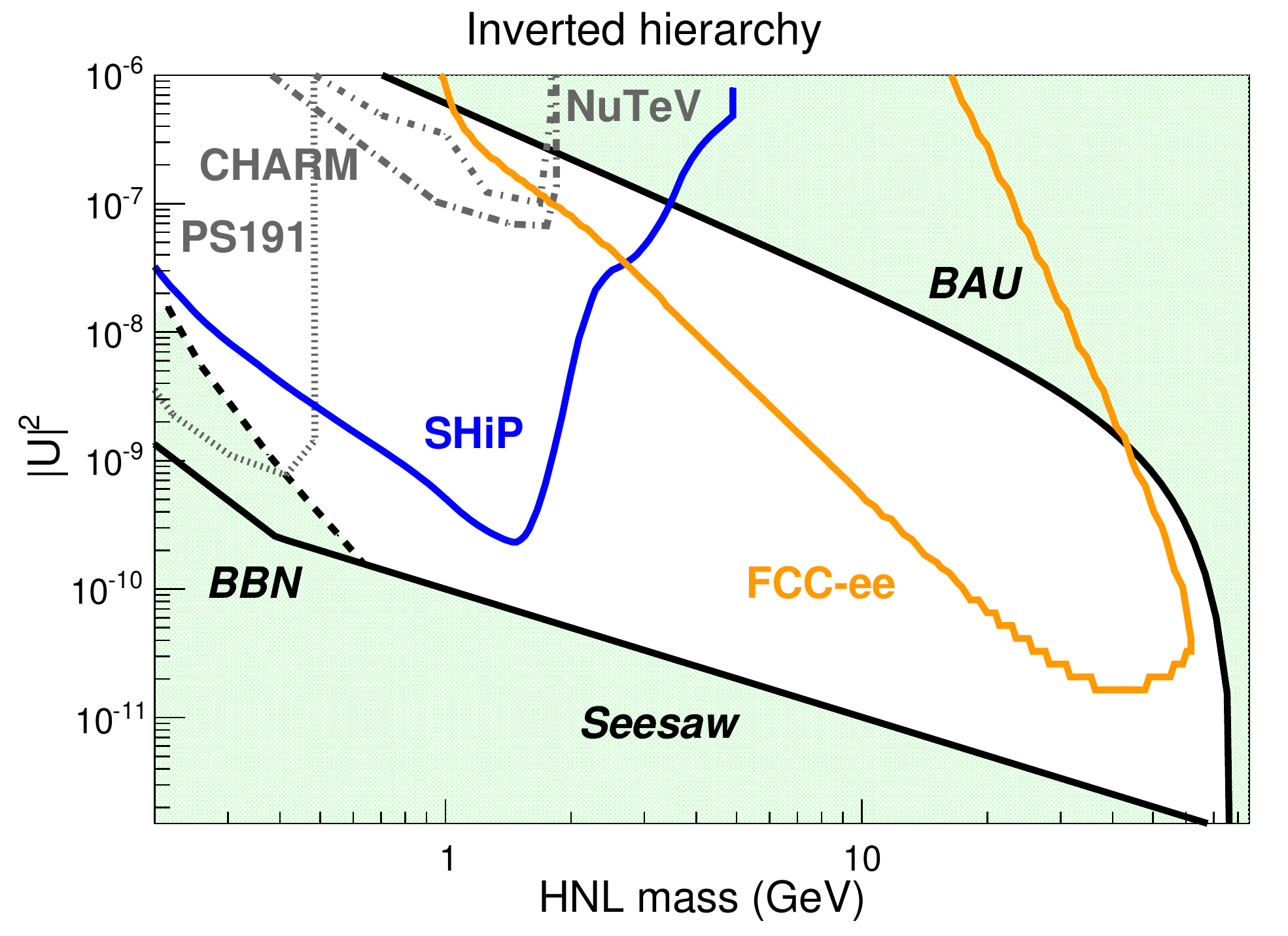}
		\caption{Decay length 10-100~cm, $10^{12}$ $Z^0$}
	\end{subfigure}
	\begin{subfigure}[]{\linewidth}
		\includegraphics[width=8.0cm]{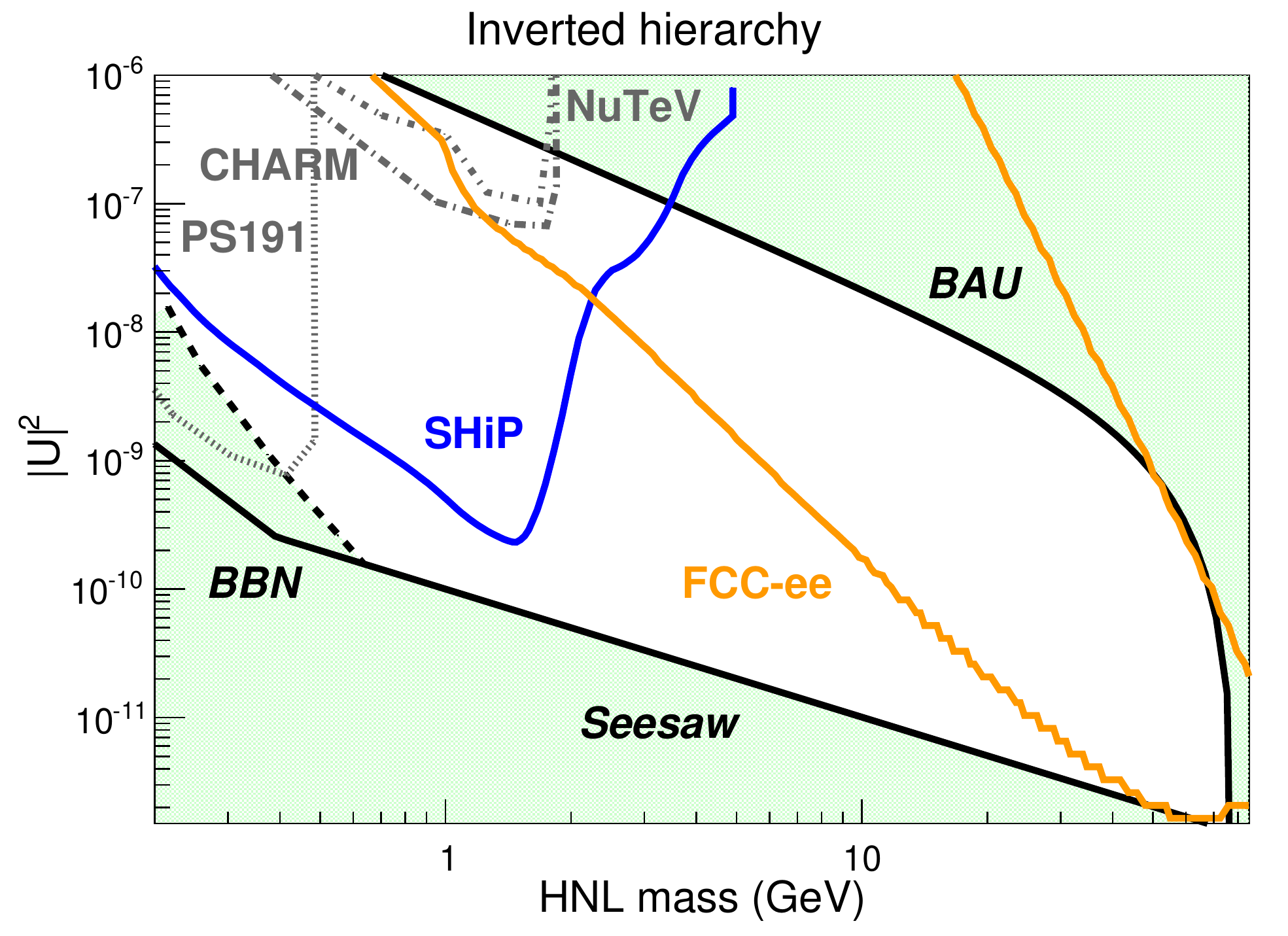}
		\caption{Decay length 10-100~cm, $10^{13}$ $Z^0$}
	\end{subfigure}
	\begin{subfigure}[]{\linewidth}
		\includegraphics[width=8.0cm]{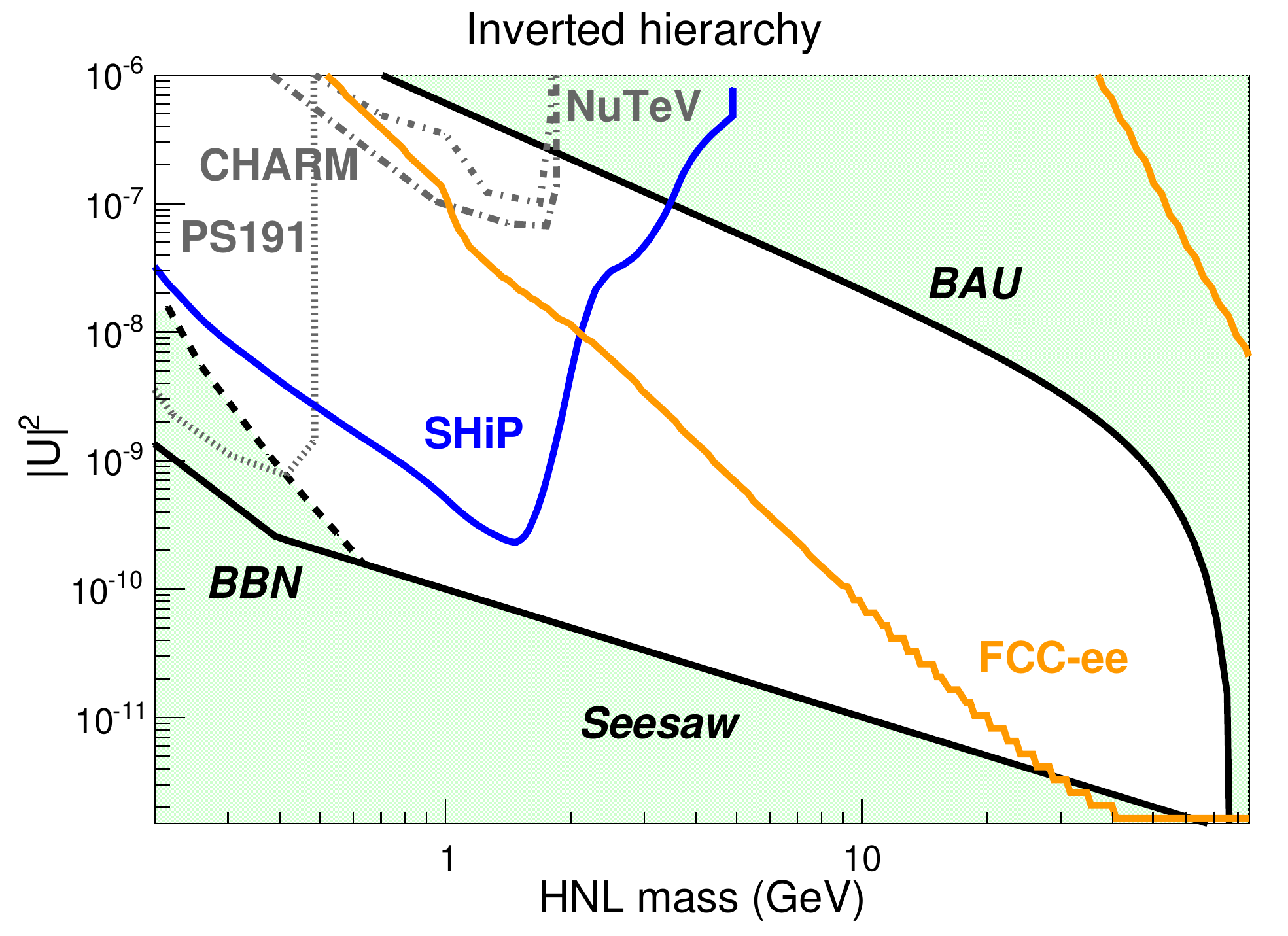}
		\caption{Decay length 0.01-500~cm, $10^{13}$ $Z^0$}
	\end{subfigure}
	\caption{Regions of sensitivity for sterile neutrinos as a function of mass and mixing to light neutrinos (inverted hierarchy): for $10^{12}$ $Z$ decays occurring between 10~cm and 1~m from the interaction point (a), same for $10^{13}$ $Z$ decays (b), for $10^{13}$ $Z$ decays occurring between 100~$\mu$m and 1~m from the interaction point (c).}\label{fig:sensitivity-ih}
\end{figure}

A first analysis of the sensitivity has been performed to evaluate the region of heavy neutrino mass and mixing in which the heavy neutrinos could be detected. So far the only requirement has been that the decay length is larger than a minimal vertex displacement and contained within a detector of given radius. Several examples are given in Figure~\ref{fig:sensitivity-nh} for the normal hierarchy and in Figure~\ref{fig:sensitivity-ih} for the more favorable case of the inverted hierarchy. It is clear that the ability to detect long decays is the most efficient way to push the sensitivity to small couplings. For a 5~m detector the full region of interest is covered for heavy neutrino masses between 30 and 80~GeV. The region of sensitivity of the proposed SHiP experiment~\cite{Bonivento:2013jag} is also shown, displaying sensitivity for masses up to the charm mass. 

\section{Conclusions}
The prospect of an $e^+e^-$ multi-Tera $Z$ factory would make the hunt for the right-handed partners of the light neutrinos an exciting and distinct possibility.
Significant work remains to be done, in order to solidly demonstrate that no unforeseen background can mimic the rather dramatic signature of a heavy neutrino decaying in the $e^+e^-$ detector. However, the preliminary studies presented here look extremely promising and should motivate further studies.

\section*{Acknowledgements}
We are indebted to all our colleagues from the TLEP/FCC-ee design study and from the SHiP collaboration for invaluable comments and encouragements.

\bibliographystyle{elsarticle-num}
\bibliography{TLEP}







\end{document}